\documentclass[dvips,showpacs,showkeys,aip,jmp,eqsecnum,floatfix]{revtex4} %
\input{epsf}
\usepackage{pstricks}
\usepackage{amsmath,amsfonts,amsthm,eucal,amssymb}

\DeclareMathOperator{\dd}{\textrm{d}}

\newcommand{\cpt}{\mathcal{CPT}}
 \newcommand{\ws}{\mathcal{W}\,}
 \newcommand{\cc}{\mathcal{C}}

 \newcommand{\pt}{\mathcal{PT}}
 \newcommand{\wm}{\mathcal{W}_{m}}
 \newcommand{\p}{\mathcal{P}}
 \newcommand{\tm}{\mathcal{T}}
 
 \newcommand{\uu}{\mathcal{U}}

\begin{document}


\title{\bf $\cpt$-conserved effective mass Hamiltonians through first and higher order
 charge operator $\cc$ in a supersymmetric framework}

\author{B.~Bagchi
 and A.~Banerjee\footnote{Permanent address: Department of Mathematics, Krishnath College, Berhampore,
 Murshidabad 742101, India}}
 \affiliation{Department of Applied Mathematics, University of Calcutta, 92
 Acharya Prafulla Chandra Road, Kolkata -- 700009, India \\ e-mail: bbagchi123@rediffmail.com, abhijit\_banerjee@hotmail.com}
 \author{A.~Ganguly}
 \affiliation{Department of  Mathematics, Indian Institute of Technology, Kharagpur 721302, India \\ e-mail: gangulyasish@rediffmail.com,
        aganguly@maths.iitkgp.ernet.in}

\date{\today}

\begin{abstract}
This paper examines the features of a generalized
position-dependent mass Hamiltonian $H_m$ in a supersymmetric
framework in which the constraints of pseudo-Hermiticity and
$\cpt$ are naturally embedded.  Different representations of the
charge operator are considered that lead to new mass-deformed
superpotentials $\wm(x)$ which are inherently $\pt$-symmetric. The
qualitative spectral behavior of $H_m$ is studied and several
interesting consequences are noted.
\end{abstract}

\keywords{Position-dependent (Effective) mass, $\cpt$-
 conservation, supersymmetric quantum mechanics, $\pt$ -symmetry, pseudo-Hermiticity}

\pacs{02.30.Tb, 03.65.Ca, 03.65.Db, 03.65.Ge}
\vspace*{.2cm}
\mbox{A version of this article will appear in Journal of
Mathematical Physics}

\maketitle

\section{Introduction}\label{intro}

Non-Hermitian systems admitting $\pt$-symmetry (i.e. invariance
under a combined action of parity $\p$ and time-reversal $\tm$)
have been a subject matter of intense interest \cite{ben1, ben2}.
$\pt$-symmetry has an interesting implication that the whole class
of Schr\"{o}dinger Hamiltonians coming under its assignment
namely, $H=p^2/2m+V(x)$ defined on the real line $x\in\mathbb{R}$,
where the potential is typically $V(x)=V^*(-x)$, may possess real
or conjugate pairs of energy eigenvalues under certain conditions
related to $\pt$ being unbroken (i.e. exact) or spontaneously
broken. It has also been realized that the concept of
$\pt$-symmetry has its roots in the theory of pseudo-Hermitian
operators and that pseudo-Hermiticity serves as one of the
plausible necessary and sufficient conditions for the reality of
the spectrum \cite{mos1}.

 In \cite{cal1} a set of intertwining relations
\begin{equation}\label{eta-pseudo}
 H\zeta = \zeta H^{\dagger}
\end{equation}
was studied in which a Hermitian operator $\zeta$ was proposed to
be expressed as a product of the charge operator $\cc$ and parity
operator $\p$
\begin{equation}\label{zeta}
\zeta= \cc \p \qquad (\zeta = \zeta^{\dagger})\, .
\end{equation}
It is straightforward to see that equations (\ref{eta-pseudo}) and
(\ref{zeta}) together imply the $\cpt$ conservation of the
Hamiltonian $H$, $\tm$ being the time reversal operator
\begin{equation}\label{cpt}
    \cpt H=H\cpt\, .
\end{equation}
Interestingly, it also follows from (\ref{eta-pseudo}) that the
operator $\zeta^{-1}$, if it exists, also fulfills the
intertwining relations
\begin{equation}\label{inv-intertwin}
 H^{\dagger}\zeta^{-1}=\zeta^{-1}H
\end{equation}
implying that $H$ is pseudo-Hermitian with respect to
$\zeta^{-1}$. This can be verified as follows:
\begin{equation}\label{verify-pseudo}
   <\psi,H\phi>_{\zeta^{-1}}=<\psi,\zeta^{-1}H\phi>=<\psi,H^{\dagger}\zeta^{-1}\phi>
=<H\psi,\zeta^{-1}\phi>=<H\psi,\phi>_{\zeta^{-1}}\, .
\end{equation}

Differential realizations for  $\cc$ have been considered in the
literature such as  $\cc = \dd/\dd x+ \ws (x)$ for the first-order
\cite{cal1} and  $\cc = \dd^2/\dd x^2+\ws (x)\dd/\dd x+ \uu_0 (x)$
for the second-order case \cite{bag1}. The aims of such models
have been to search for closed-form solutions and to work out the
solvability criterion of the embedded Hamiltonian.

In this article we intend to investigate these and related aspects
of pseudo-Hermiticity and $\cpt$-conservation for extended
versions of Schr\"odinger equation admitting SUSY in a
position-dependent (effective) mass (PDM) framework. The 1-D
effective mass Hamiltonian $H\rightarrow H_m$ obeys (in the atomic
unit defined by $\hbar^{2}=2$) in real spatial coordinate
\cite{bag}:
\begin{equation}\label{emsc}
 H_{m}(x)\psi_n(x) \equiv \left (-\partial\left [\frac{1}{m(x)}\partial\right ]
  + \widetilde{V}_{\textrm{m}}(x)\right )\psi_n(x)
 = E_n \psi_n(x), \qquad \widetilde{V}_{\textrm{m}}(x)=V_{\textrm{m}}(x)+\rho(m)\, ,
\end{equation}
where $m(x)$ is a real valued mass function in the presence of a
complex potential $\widetilde{V}_{\textrm{m}}(x)$:
\begin{equation}\label{complex-potential}
 V_{\textrm{m}}(x)= V^R_{\textrm{m}}(x)+\textrm{i}V^I_{\textrm{m}}(x)\, .
\end{equation}
 In equation~(\ref{emsc}), the  mass-dependent function $\rho(m)$ has the
 form \cite{vonRoos}
\begin{equation}\label{rho}
\rho(m)=\frac{1+b}{2}\frac{m''(x)}{m^{2}(x)}-c\frac{{m'}^{2}(x)}{m^{3}(x)},
    \qquad c=1+b+a(a+b+1),
\end{equation}
where $a \mbox{ and } b$ are the usual ambiguity parameters
\cite{mus} typical to the effective-mass models. Position
dependence in mass shows up in different areas of physics -
semiconductors \cite{basbk}, quantum dots \cite{ser} , $^3$He
clusters \cite{bar} and many
 more.  A number of papers have been written on the issue of PDM in this rapidly expanding
 literature \cite{jon,bag6,q4,cap,koc2,alh,koc3,q1,ou,yu,zno,levai2,bag3,as10,q3,as9,ag1,roy,yesi,DSDA,RR,TAN,BBTAN}.

Note that the question of boundedness and invertibility of the
operator $\zeta$, assuming an explicit representation for it was
addressed in \cite{cal1} for the constant-mass case. In the PDM
scenario, the problem is trickier and will be taken up elsewhere.

\section{Pseudo-Hermiticity and $\cpt$-symmetry in a supersymmetric framework}\label{pseudo-cpt-setup}

In the framework of supersymmetric (SUSY) quantum mechanics
 \cite{Cooper,Junker,bagbook,levai1},
an underlying  anticommutator $K$ of the supercharges  $Q$ and
$\overline{Q}$ can be explicitly constructed by specifying the
following representation
\begin{equation}\label{sup=alg}
    K=\{ Q, \overline{Q} \} =\left (
                       \begin{array}{cc}
                             \zeta\zeta^* & 0 \\
                      0 & \zeta^*\zeta \\
                        \end{array}
\right )\, ,
\end{equation}
where $Q$ and $\overline{Q}$ are defined in terms of the operator
$\zeta$ and its complex conjugate $\zeta^*$
\begin{equation}\label{supe-chrg}
 Q=\left (
                       \begin{array}{cc}
                             0 & \zeta \\
                       0 & 0 \\
                        \end{array}
\right )\, , \quad
 \overline{Q}=\left (
                          \begin{array}{cc}
                           0 & 0 \\
                           \zeta^* & 0 \\
                        \end{array}
\right )\, .
\end{equation}

Assuming polynomial expansions
\begin{equation}\label{polynomial-identity}
 \zeta\zeta^*=\sum_{k=0}^{N} l_k\left (H_m\right )^{N-k}\, , \quad \zeta^*\zeta=\sum_{k=0}^{N} l_k\left ( H^*_m\right )^{N-k}\, ,
 \qquad [l_0\equiv 1\, ,   H_m^0\equiv I_2]
\end{equation}
we get by post-multiplying the first relation and pre-multiplying
the second relation above by $\zeta$ and subtracting
\begin{equation}\label{step1}
    0=\sum_{k=0}^{N-1}\left [ l_k\left (H_m\right
    )^{N-k}\zeta\right ]-\sum_{k=0}^{N-1}\left [ l_k\zeta\left ( H^*_m\right
    )^{N-k}\right ]\, .
\end{equation}
Similarly, by pre-multiplying the first relation and
post-multiplying the second relation by $\zeta^*$ and subtracting
 \begin{equation}\label{step2}
    0=\sum_{k=0}^{N-1}\left [ l_k\zeta^*\left (H_m\right
    )^{N-k}\right ]-\sum_{k=0}^{N-1}\left [ l_k\left ( H^*_m\right
    )^{N-k}\zeta^*\right ]\, .
\end{equation}
(\ref{step1}) and (\ref{step2}) lead to the intertwining relations
\begin{equation}\label{susy-intertwining}
    H_m\zeta=\zeta H^*_m\, ,\quad  H^*_m\zeta^*=\zeta^*H_m\, .
\end{equation}
At play are also the following constraints
\begin{eqnarray}
 \mbox{Pseudo-hermiticity constraint: }  \zeta^{\dagger}=\zeta &\Rightarrow & \cc^{\dagger}[-x]=\cc[x]\, , \label{pseudo-hermiticity-constraint} \\
 \mbox{$\cpt$ constraint: }  \cpt H_m=H_m\cpt &\Rightarrow & \cc[x]H^*_m[-x]= H_m[x]\cc[x]\, ,\label{cpt-constraint-2}\\
 \mbox{SUSY constraint: } \zeta\zeta^*=\sum_{k=0}^{N} l_k H_m^{N-k}& \Rightarrow & \cc[x]\cc^*[-x]=\sum_{k=0}^{N} l_k H_m^{N-k}\, .\label{susy-polynomial1}
\end{eqnarray}

Finally, in the context of the $N$-th order SUSY, a mass-deformed
superpotential $\wm(x)$
 can be introduced which is given by the form \cite{ag1}
\begin{equation}\label{general-mass-superpotential}
 \wm(x)=\ws(x)-\frac{N}{2}\left [
 \frac{1}{\sqrt{m^{N}(x)}}\right ]'\, .
\end{equation}
where $\ws(x)$ corresponds to the superpotential of the constant
mass case. A natural consequence of
(\ref{general-mass-superpotential}) is that unlike $\ws(x)$ as in
the constant-mass case $\wm(x)$ turns out to be $\pt$-symmetric
from the pseudo-Hermiticity constraint
(\ref{pseudo-hermiticity-constraint}) as will be revealed below.

\section{First order charge operator}\label{k=1}

The first order representation of the charge operator $\cc$ in a
PDM scheme is given by
\begin{equation}\label{c1}
 \cc=\frac{1}{\sqrt{m(x)}}\frac{\dd}{\dd x}+\ws(x)\, .
\end{equation}
From (\ref{polynomial-identity}) we have for $N=1$ the projections
\begin{equation}\label{linear-susy}
    \zeta\zeta^*=H_m+l_1\, , \quad \zeta^*\zeta=H^*_m+l_1\, .
\end{equation}

Imposing the pseudo-hermiticity restriction
(\ref{pseudo-hermiticity-constraint}), we have the solutions:
\begin{equation}\label{pseudo-requirement-k1}
    m(x)=m(-x)\, , \quad    \ws(x)=\ws^*(-x)-\frac{1}{2}\frac{m'(x)}{m^{3/2}(x)}\, .
\end{equation}

It is evident from (\ref{general-mass-superpotential}) and
(\ref{pseudo-requirement-k1}) that
\begin{equation}\label{pseuodo-constraint-k1}
 \wm(x)=\pt\, \wm(x)\, .
\end{equation}
implying $\wm(x)$ to be $\pt$-symmetric and the mass function to
be parity-invariant. As remarked earlier, $\ws(x)$ ceases to be
$\pt$-symmetric. A $\pt$-symmetric $\wm(x)$ can be implemented by
choosing for $\ws(x)$ the form say, $\ws(x)= \exp(i\alpha x) +
h(x)$ where $\alpha \in \mathbb{R}$ and a non-$\pt$ $h(x)$ can be
confronted with a suitable parity-invariant mass function leaving
$\wm(x)$ to be $\pt$-symmetric. The following concrete example is
one we have in mind
\begin{equation}\label{exam-1st}
\ws(x)=\exp(i\alpha x)-\sin (x),\quad
m(x)=\frac{1}{4}\sec^{2}(x)\quad \Rightarrow \wm (x)=\exp(i\alpha
x) \quad 0<x<\frac{\pi}{2}
\end{equation}
where $\wm (x)$ is a periodic potential.

Turning to the $\cpt$-constraint (\ref{cpt-constraint-2}) and
using (\ref{c1}) we get two relations. While comparison of the
$\partial$-term yields the difference
\begin{equation}\label{non-pt-symmtery-pot-k1}
 \widetilde{V}_{\textrm{m}}(x)-\pt
    \widetilde{V}_{\textrm{m}}(x)=2\frac{\wm'(x)}{\sqrt{m(x)}}\, ,
\end{equation}
the remaining part results in a first-order differential equation
which can be readily integrated to provide for
$\widetilde{V}_{\textrm{m}}(x)$ the expression
\begin{equation}\label{pot-k1-first}
    \widetilde{V}_{\textrm{m}}(x)=\wm^2(x)+\frac{\wm'(x)}{\sqrt{m(x)}}+\frac{1}{4}\frac{m''}{m^2}-\frac{7}{16}\frac{{m'}^2}{m^3}+\Lambda\, ,
\end{equation}
where $\Lambda$ is an arbitrary constant of integration.

A non-trivial form for $\widetilde{V}_{\textrm{m}}(x)$ is also
obtained on employing the SUSY constraint (\ref{susy-polynomial1})
for $N=1$ namely
\begin{equation}\label{susy-1}
    \cc[x]\cc^*[-x]=H_m+l_1\, .
\end{equation}

A comparison between (\ref{pot-k1-first}) and (\ref{susy-1}) fixes
$\Lambda =-l_1$. (\ref{susy-1}) is our final form of
$\widetilde{V}_{\textrm{m}}(x)$ for the $N=1$ case. Note that the
underlying $\cpt$-invariance  has the implication
\begin{equation}\label{cpt-invariant}
H_{m}\cpt \psi(x)=\cpt H_{m}\psi(x)=\cpt E\psi(x)=E^* \cpt
\psi(x).
\end{equation}
Thus if  $(\psi,E)$ is an eigenpair of a $\cpt$-invariant PDM
Hamiltonian $H_m$, then $(\cpt\psi,E^*)$ must form another
eigenpair provided $\cpt\psi\neq 0$. Thus $\cpt$-invariance of
$\psi$ leads to the corresponding Hamiltonian having real
eigenvalues.

From the first relation of (\ref{polynomial-identity}) and
(\ref{emsc}), the ground state $\psi_0$ in the $N=1$ case has to
obey
\begin{equation}\label{zero-mode-1storder}
    \zeta^*\psi_0 (x)=0\Rightarrow \left [ \frac{1}{\sqrt{m(x)}}\partial +\ws^*(x)\right ]\psi_0 (-x)=0\, ,
\end{equation}
with the lowest eigenvalue $-l_1$. From (\ref{zero-mode-1storder})
we find
\begin{equation}\label{sol-k1}
    \psi_0(x)=\mathcal{N}_0m^{1/4}(x)\exp \left [ \int^x\sqrt{m(y)}\wm (y)\dd y\right ]\, ,
\end{equation}
 $\mathcal{N}_0$ being the normalization constant.  Note that $\psi_0(x)$
is non-$\pt$-symmetric.

\section{Second order charge operator}\label{k=2}

We now look at the following mass-dependent second order
representation of the charge operator $\cc$
\begin{equation}\label{c2}
 \cc=\frac{1}{m(x)}\frac{\dd^2}{\dd x^2}+\ws(x)\frac{\dd}{\dd x}+\uu_0(x)\, ,
 \end{equation}
 accompanied by the $N=2$ SUSY representations
 \begin{equation}\label{polynomial-id-k2}
 \zeta\zeta^* =H_m^2+l_1H_m+l_2I_2 \, .
\end{equation}
as follows from (\ref{polynomial-identity}). For the literature on
$N=2$ SUSY in the constant-mass case we refer to the readers
\cite{andrianov1,andrianov2,bagchi,andrianov4,correa,fernandez}.

Employing the pseudo-Hermiticity requirement
(\ref{pseudo-hermiticity-constraint}) gives the following
solutions
\begin{equation}\label{common-prop-k2}
    m(-x)=m(x)\, , \qquad \wm(x)=\pt\,\wm(x)\,
\end{equation}
which are similar to the $N=1$ case  i.e. $\wm(x)$ is
$\pt$-symmetric and the mass function $m(x)$ is parity-invariant.
Note that according to (\ref{general-mass-superpotential}),
$\wm(x)$ is related to the constant-mass superpotential $\ws(x)$
by
 \begin{equation}\label{mass-deform-sup-k2}
    \wm(x)=\ws(x)-\left (\frac{1}{m}\right )'\, .
\end{equation}
As an illustrative example we can take this time
\begin{equation}\label{exam-2nd}
\ws (x)=\exp(i\alpha x)-\sin (x), \quad m(x)=\sec (x) \qquad
\alpha \in \mathbb{R},\quad 0<x<\frac{\pi}{2}
\end{equation}
leading again to a periodic $\pt$-symmetric $\wm (x)=\exp(i\alpha
x)$.

Apart from (\ref{common-prop-k2}), the pseudo-Hermiticity
condition also furnishes another relation namely
\begin{equation}\label{pt-asym-u0-k2}
    \triangle\,\uu_0(x)\equiv\uu_0(x)-\pt\, \uu_0(x)=\wm'(x)\, ,
\end{equation}
which reflects the non-$\pt$-symmetric character of the function
$\uu_0 (x)$ present in (\ref{c2}).

Next, consideration of the $\cpt$ requirement
(\ref{cpt-constraint-2}) furnishes
\begin{equation}\label{pt-asym-v-k2}
    \triangle \widetilde{V}_{\textrm{m}}(x)\equiv \widetilde{V}_{\textrm{m}}(x)-\pt \widetilde{V}_{\textrm{m}}(x)=2\wm'(x)+\frac{m'}{m}\wm(x)\,
\end{equation}
which is slightly different in form from the $N=1$ result
(\ref{pot-k1-first}). In (\ref{pt-asym-v-k2})
 $\widetilde{V}_{\textrm{m}}(x)$ is restricted by
\begin{equation}\label{v-k2}
  \widetilde{V}_{\textrm{m}}(x)=\triangle \widetilde{V}_{\textrm{m}}(x)+f(x)-\uu_0(x)+\Lambda\,
\end{equation}
where
\begin{equation}\label{f}
f(x)=\frac{1}{2}\left [ m\wm^2(x)-\frac{m'}{m}\wm(x)-\wm'(x)
\right ]\,
\end{equation}
on making use of (\ref{common-prop-k2}) and (\ref{pt-asym-u0-k2}).
The constant $\Lambda$ appears in (\ref{v-k2}) through the process
of integration and is left arbitrary at this stage.

In addition to (\ref{pt-asym-v-k2}) and (\ref{v-k2}), the
non-$\pt$-function $\uu_0(x)$ has to satisfy the differential
equation
\begin{equation}\label{undressed-u-k2}
    \left [ \,\frac{\uu'_0(x)}{m(x)}\,\right ]'-\uu_0(x)\triangle
    \widetilde{V}_{\textrm{m}}(x)+\wm(x)\left \{ \pt\, \widetilde{V}_{\textrm{m}}(x)\right \}'
    +\left [\, \frac{\left \{ \pt\, \widetilde{V}_{\textrm{m}}(x)\right \}'}{m(x)}\, \right ]'=0\, .
\end{equation}

Substitution of (\ref{v-k2}) into (\ref{undressed-u-k2}) converts
it to the form
 \begin{equation}\label{dressed-u-k2}
    \uu'_0(x)\wm(x)+\uu_0(x)\left [2\wm'(x)+\frac{m'}{m}\wm(x)\right ]
    =\frac{f''(x)}{m}+f'(x)\left [\wm(x)-\frac{m'}{m^2}\right ]\, ,
\end{equation}
which may be solved to arrive at
\begin{equation}\label{intermediate-u-k2}
    \uu_0(x)=\frac{f'(x)}{m(x)\wm(x)}+\frac{f^2(x)}{m(x)\wm^2(x)}+\frac{\Theta}{m\wm^2(x)}\, .
\end{equation}
$\Theta$ being an arbitrary constant of integration.

We now attend to the SUSY constraint (\ref{susy-polynomial1}).
Here we need to compare the five coefficients of
 $\partial^{\ell}\, ,\ell=0,1,2,3,4$. While the first two produce solutions similar to (\ref{common-prop-k2}), the last three respectively
 yields the following three relations:
 \begin{eqnarray}
    2\frac{1}{m(x)} \widetilde{V}_{\textrm{m}}(x)+\frac{1}{m(x)}\left [\uu_0(x)+\pt\, \uu_0(x)\right ]
      -2\frac{1}{m(x)} \wm'(x)&=&\wm^2(x)-(\frac{1}{m(x)})'\wm(x)-l_1\frac{1}{m(x)}\, ,\label{1st-susy-relation-k2}\\
    \frac{1}{m(x)}\left [\triangle\uu_0(x)\right
    ]'+\wm(x)\triangle\uu_0(x)&=&\frac{1}{m(x)} \wm''(x)+\left
    [\wm^2(x)/2\right ]'\, ,\label{2nd-susy-relation-k2}\\
    \left [\frac{1}{m(x)} \left \{\pt\,\uu_0(x)\right \}'\,\right
    ]'+\wm(x)\left \{\pt\,\uu_0(x)\right
    \}'+\uu_0(x)\pt\,\uu_0(x)&=&\widetilde{V}_{\textrm{m}}^2(x)+l_1\widetilde{V}_{\textrm{m}}(x)+l_2-\left [\frac{1}{m(x)}
    \widetilde{V}'_{\textrm{m}}(x)\right ]'\, .\label{3rd-susy-relation-k2}
\end{eqnarray}

To tackle the set of equations
(\ref{1st-susy-relation-k2})--(\ref{3rd-susy-relation-k2}), we
observe that the second equation here can be integrated out
entirely to have
\begin{equation}\label{t-gen-k2}
    \triangle\uu_0(x)=\wm'(x)+\textrm{C}\exp \left [-\int^x
    m(y)\wm(y)\dd y\right ]\, ,
\end{equation}
where $\textrm{C}$ is a constant of integration. But $\textrm{C}$
has to be set equal to zero to be consistent with
(\ref{pt-asym-u0-k2}). So we are left with (\ref{pt-asym-u0-k2})
only. Incorporating it along with (\ref{pt-asym-v-k2}),(\ref{f})
and (\ref{1st-susy-relation-k2}) $\widetilde{V}_{\textrm{m}}(x)$
reads
\begin{equation}\label{expn-susy-pot-k2}
    \widetilde{V}_{\textrm{m}}(x)=\triangle \widetilde{V}_{\textrm{m}}(x)+f(x)-\uu_0(x)-l_1/2\, ,
\end{equation}

 Then looking at (\ref{v-k2}) prompts us to identify $\Lambda=-\frac{l_1}{2}$ and recast $\widetilde{V}_{\textrm{m}}(x)$ as
 \begin{equation}\label{assignment2-k2}
        \widetilde{V}_{\textrm{m}}(x)=\frac{3}{2}\wm'(x)+\frac{m'}{2m}\wm(x)+\frac{m}{2}\wm^2(x)-\uu_0(x)-\frac{l_1}{2}\, .
\end{equation}

 We now focus on the remaining SUSY constraint (\ref{3rd-susy-relation-k2}). This can be converted to a second-order differential equation
  \begin{equation}\label{last-susy-k2}
    \left [ \frac{1}{m(x)}\left \{ \uu_0(x)+\widetilde{V}^*_{\textrm{m}}(-x)\right \}'\right ]'
    -\uu'_0(x)\wm(x)+\left [\, \uu_0(x)-\wm'(x)\,\right ]\uu_0(x)
     =\left [ \widetilde{V}^*_{\textrm{m}}(-x)+\frac{l_1}{2}\right ]^2+\left ( l_2-\frac{l_1^2}{4}\right )\, ,
\end{equation}
by applying the $\pt$-operator on both sides and rearranging. Note
that the action of $\pt$ on any function $g(x)$ is to be
understood in the usual sense: $\pt\,
g(x)=g^*(-x)\,,\pt\,g'(x)=-g^{*'}(-x)$ and so on. The nonlinear
term $\uu_0^2$ in (\ref{last-susy-k2}) is redundant and can be
eliminated in the following way. Using the relation
$\widetilde{V}^*_{\textrm{m}}(-x)\equiv \pt\,
\widetilde{V}_{\textrm{m}}(x)=\widetilde{V}_{\textrm{m}}(x)-\triangle
\widetilde{V}_{\textrm{m}}(x)$,  (\ref{assignment2-k2}) results in
\begin{equation}\label{alternate-forms}
    \widetilde{V}^*_{\textrm{m}}(-x)+\uu_0(x)=f(x)-\frac{l_1}{2}\,  .
\end{equation}

Employing (\ref{alternate-forms}), (\ref{last-susy-k2}) can be
reduced to the first order form
\begin{equation}\label{alternate-u0-k2}
    \wm(x)\uu'_0(x)+\left [\, \wm'(x)-2f(x)\,\right ]\uu_0(x)
     =\left [ \frac{1}{m(x)}(x)f'(x)\right ]'-f^2(x)+\left [ \frac{l_1^2}{4}-l_2\right ].
\end{equation}
Equation (\ref{alternate-u0-k2}) which essentially results from
the SUSY constraint (\ref{3rd-susy-relation-k2}) is consistent
with the $\cpt$ equation
 (\ref{dressed-u-k2}) for $\uu_0(x)$ given by (\ref{intermediate-u-k2})
  should we identify $\Theta =\left [ l_2-\frac{l_1^2}{4}\right ]$. In terms of $\delta=+\sqrt{l_1^2-4l_2}$ we express $\uu_0 (x)$ as
\begin{eqnarray}\label{assignment3-final-u-k2}
    \uu_0 (x) &=& \frac{m(x)\wm^2 (x)}{4}+\frac{\wm'(x)}{2}-\frac{\wm''(x)}{2m(x)\wm(x)}
     +\frac{1}{m(x)}\left (\frac{\wm'(x)}{2\wm(x)}\right )^2\nonumber \\
     && +\frac{3}{4}\frac{m^{'^2}(x)}{m^3 (x)}-\frac{m''(x)}{2m^2 (x)}
     -\frac{1}{m(x)}\left(\frac{\delta}{2\wm(x)}\right )^2\, .
\end{eqnarray}
As a specific example we can go for the choice (\ref{exam-2nd})
which would give
\begin{eqnarray}\label{exam-u-2nd}
\uu_0 (x) &=& \frac{1}{4}\sec(x) \exp(2i\alpha x)-\frac{{\delta}^2}{4}\cos(x) \exp(-2i\alpha x)+\frac{i\alpha}{2}\exp(i\alpha x)\nonumber\\
&& +\frac{\alpha^2}{4}\cos(x)+\frac{1}{4}\sin^2
(x)\sec(x)-\frac{1}{2}\sec(x).
\end{eqnarray}
Evidently $\uu_0 (x)$ is non-$\pt$-symmetric.

Let us now analyze the solution of the zero-mode equation
\begin{equation}\label{zero-mode-2ndorder}
    \zeta^*\psi(x)=0\Rightarrow \left [
    \frac{1}{m}\partial^2+\ws^*(x)\partial +\uu^*_0(x)\right ]
    \psi(-x)=0\, .
\end{equation}
Two linearly independent solutions of zero-mode equation
(\ref{zero-mode-2ndorder}) may be expressed in the following
compact form (see for details \cite{ag1}):
\begin{equation}\label{lowest-state-2nd-order}
    \psi_j(x)=\mathcal{N}_j\sqrt{m(x)\wm(x)}\exp \left [
    \int^xF_j(y)\dd y\right ]\, ,
\end{equation}
where
\begin{equation}\label{F}
    F_j(x)=\frac{m(x)\wm^2(x)+(-1)^j\delta}{2\wm(x)}\, , \quad j=1,2.
\end{equation}
These solutions will correspond the ground and first excited
states of $H_m$, a feature known in the quadratic SUSY algebra.

Now it follows from the quadratic SUSY algebra
(\ref{polynomial-id-k2}) that the  lowest eigenvalues of $H_m$ are
roots of the following quadratic equation
\begin{equation}\label{eigenvalue}
    E^2+l_1E+l_2=0\, \Rrightarrow E_0=-\frac{l_1+\delta}{2}\, ,
    E_1=-\frac{l_1-\delta}{2}\, .
\end{equation}
It is clear that the lowest two eigenvalues $E_{0,1}$\r will be
purely real if and only if the SUSY constants $l_1,l_2$ satisfy
following inequality
\begin{equation}\label{inequality}
    l_1^2\geq 4l_2\, .
\end{equation}
It may be pointed out that the condition $l_1^2\geq 4l_2$ was
identified with the
 reducibility of the second-order SUSY construction \cite{andrianov2} in the
context of Hermitian QM. In non-Hermitian
 QM, we have shown that the same condition is related with the reality of the spectra.

\section{$N$-th order charge operator}\label{general-cc}

The charge conjugate operator $\cc$ may be represented as $N$-th
order differential operator with $N$ coefficient functions
\begin{equation}\label{general-c}
    \cc=\frac{1}{\sqrt{m^{N}}}\partial^{N}+\ws(x)\partial^{N -1}
     +\sum_{j=0}^{N-2}\uu_j(x)\partial^j\, , \quad N=1,2,\ldots.
\end{equation}

Some of the previous results are possible to generalize. Firstly,
the pseudo-Hermiticity constraint
(\ref{pseudo-hermiticity-constraint}) need to be compared order by
order from both sides for $N$-th order representation
(\ref{general-c}) of $\cc$. To do this, we note that the
contributions from the
 adjoint operation on the term $g(x)\partial^{\ell}$
 may be computed
 using the Libneitz rule as follows
 \begin{equation}\label{libneitz}
    \left [ g(x)\partial^{\ell}\right ]^{\dagger}
      =(-1)^{\ell}\partial^{\ell}\left [ g(x) \right ]
       =(-1)^{\ell}\sum_{r=0}^{\ell}
       \left [ ^{\ell}\textrm{C}_r\partial^r\{ g(x)\}\partial^{\ell-r}\right ]\, .
\end{equation}

Then order by order comparison gives the following restrictions on
the coefficient functions in the charge operator $\cc$ given by
(\ref{general-c})
\begin{equation}\label{m-even-sup-pt}
\left . \begin{array}{lll}
 \ell=N: & \p \left [ m(x)\right ]=m(x)\, ,& (\mbox{mass is parity-invariant}) \\
 \ell=N-1: & \pt\left [ \wm(x)\right ]=\wm(x)\, . & (\mbox{ superpotential is }\pt\mbox{-invariant})
 \end{array} \right \}\, ,\quad N=1,2,3,\ldots\:\: .
\end{equation}
One may compare the general result derived above with the
 corresponding results for $N=1$ [ see (\ref{pseudo-requirement-k1}) and (\ref{pseuodo-constraint-k1})]
 and for $N=2$ [ see (\ref{common-prop-k2})]. Note that the
second condition means as usual that the $\mbox{Re}\,\wm(x)$ is an
even function while its imaginary part
 $\mbox{Im}\, \wm(x)$ is an odd function.

 In contrast to the superpotential
$\wm$, the functions $\uu_{j},j=0 \mbox{ to }N-2\, ,$ are not
$\pt$-symmetric. For instance, for $\ell=N-2$ and $\ell=N-3$ we
have
\begin{equation}\label{k-2-1}
 \hspace{-7cm}\uu_{N-2}(x)-\pt \left [
    \uu_{N-2}(x)\right ]=(N-1)\wm'(x)\,
    ,\quad N=2,3,\ldots
\end{equation}
\begin{equation}\label{k-3-1}
    \uu_{N-3}(x)-\pt \left [
    \uu_{N-3}(x)\right ]=(N-2)\left [ -\frac{N-1}{2}\left \{ \frac{N}{6}\left (\frac{1}{\sqrt{m^{N}}}\right )'''
      +\wm''\right \}+\uu'_{N-2}(x)\right ]\,
      ,\quad N=3,4,\ldots\, ,
\end{equation}
and so on. More generally,
\begin{equation}\label{gen1}
    \triangle \uu_{N-s}(x)= \uu_{N-s}(x) - \pt \uu_{N-s}(x)
    \hspace{1pt}^{N}\textrm{C}_s\partial^s\left (\frac{1}{\sqrt{m^{N}}}\right )+
      \hspace{1pt}^{N-1}\textrm{C}_{s-1}\partial^{s-1}
      \left [ \pt \ws(x)\right ]+\sum_{j=1}^{s-2}\hspace{1pt}^{N-s+j}\textrm{C}_j\partial^j
       \left [ \pt \{ \uu_{N-s+j}(x)\}\right ]\, ,
\end{equation}
From the results, it is clear that the pseudo-Hermiticity
constraints measure the amount of $\pt$-asymmetry in the
coefficient functions. In particular, the measure is zero for
first coefficient $m(x)$ and mass-deformed superpotential
$\wm(x)$.

Next comparing the coefficients of each derivative
$\partial^{\ell}$ for $\ell=0,1,\ldots N+2$ from both sides of the
$\cpt$-constraint (\ref{cpt-constraint-2}), a straightforward
calculation shows
\begin{equation}\label{pt-asymmetry-pot}
    \triangle \widetilde{V}_{\textrm{m}}(x)\equiv \widetilde{V}_{\textrm{m}}(x)-\pt
    \left [ \widetilde{V}_{\textrm{m}}(x)\right ]=\frac{\sqrt{m^{N}}}{m^2}\left [
    2m\wm'(x)+(N-1)m'\wm(x)\right ]\, , N=1,2,3,\ldots\:.
\end{equation}

Comparison for $\ell=N-1$ gives a closed expression for the
potential due to the integrability of the equation
\begin{eqnarray}\label{gen-pot}
    N \widetilde{V}_{\textrm{m}}(x)&=&\frac{\sqrt{m^{N}}}{m^2}\left [
    \hspace{1pt}^{N}\textrm{C}_2m'\wm(x)+m\left
    \{\sqrt{m^{N}}\wm^{^2}(x)
     +(2N-1)\wm'(x)-2\uu_{N-2}(x)\right
     \}\right ]\nonumber \\
     &&\mbox{}+\frac{N(N-2)}{48}\left [ 4(2N+1)\left
     (\frac{1}{m}\right )''+3N(N-2)m\left (\frac{1}{m}\right
     )^{'^2}\right ]+\Lambda \, ,\,N=1,2,3,\ldots\:\:,
\end{eqnarray}
where we set a convention that $\hspace{1pt}^{N}\textrm{C}_j\equiv
0\, , \uu_{N-j}\equiv 0$ for $N<j$.  Continuing this comparison up
to the term $\partial^0$, we find that for all order $N$, only two
 coefficient functions in the representation of charge operator
 $\cc$ remain independent, which are the mass function $m(x)$ and the
 superpotential $\wm$. As for instance, comparing
 $\partial^{N-2}$ from both sides of (\ref{cpt-constraint-2}),
 one obtains for $N\geq 3$
 \begin{eqnarray}
    \left [ \frac{\uu'_{N-2}(x)}{m}\right ]'-\left [ \triangle \widetilde{V}_{\textrm{m}}(x)+\hspace{1pt}^{N-1}\textrm{C}_2\left
    (\frac{1}{m}\right )''\right ]\uu_{N-2}(x)+\frac{2}{m} \uu'_{N-3}(x)
    -(N-3)\left (\frac{1}{m}\right )'\uu_{N-3}(x) &&\nonumber\\
  \mbox{}=\hspace{1pt}^{N+1}\textrm{C}_4\frac{1}{\sqrt{m^{N}}}
   \left ( \frac{1}{m}\right )^{IV}-\hspace{1pt}^{N}\textrm{C}_2 \frac{(\widetilde{V}^*_{\textrm{m}})''(-x)}{\sqrt{m^{N}}}
   +\ws(x)\left [ \hspace{1pt}^{N}\textrm{C}_3
  \left ( \frac{1}{m}\right )'''-(N-1)(\widetilde{V}^*_{\textrm{m}})'(-x)\right ]\, ,\: N=2,3,\ldots\: , &&
  \end{eqnarray}

  Similar to the first and second
order cases, a general nonlinear SUSY algebra can be set up. The
energy in such an algebra are zeros of the same $N$-th degree
polynomial
\begin{equation}\label{energy-N-order}
    E^N+l_1E^{N-1}+l_2E^{N-2}+\cdots +l_{N-2}E^2+l_{N-1}E+l_N=0\,
\end{equation}
from which we conclude that for an odd-order charge operator, the
Hamiltonian $H_m$ possesses at least one real energy eigenvalue.

 \section{Conclusion}\label{conclusion}

In this article we have studied a generalized PDM  Schr\"odinger
equation in a non-Hermitian framework. We have proposed new
differential realization for the charge operator and sought for
the solvability of the model. Several interesting consequences due
to PDM and non-Hemiticity of the Hamitonian are derived. It should
be noted that not all the results of the constant-mass
non-Hermitian system are carried over to the PDM case. In
constant-mass case, we showed that the superpotential $\ws(x)$ had
to be $\pt$-symmetric to preserve $\cpt$-symmetry and
pseudo-hermiticity. In contrast, in the present work we have shown
that the superpotential $\ws(x)$ loses its $\pt$-symmetric
property. Instead a new mass-deformed superpotential $\wm(x)$ can
be defined which turns out to be  $\pt$-symmetric. Our work
uncovers a new class of potentials $\widetilde{V}_m(x)$ admitting
$\cpt$-symmetry in PDM non-Hermitian systems. We have also
obtained extension of some of our results to a general $N$-th
order charge operator wherein the mass function remains even and
the mass-deformed superpotential  $\pt$-symmetric.

\end{document}